# Regenerative Soot-IV: The state of the carbon vapour in the regenerative sooting discharge

Shoaib Ahmad[1],[*]

[1]*National Centre for Physics, Quaid-i-Azam University Campus, Shahdara Valley, Islamabad, 44000, Pakistan*

[*]Email: sahmad.ncp@gmail.com

**Abstract.** Formation of the regenerative soot is the process of recycling and introduction of the cathode deposited carbon clusters into the discharge. The agglomerates of carbon clusters on the cathode release their constituents into the plasma that goes from the pure sputtering mode to the sooting one. The process of the regeneration of the soot that emits large carbon clusters is discussed by evaluating the state of the carbon vapour by using the characteristic line emissions from the discharge.

z

# 1 Introduction

Soot formation in a cusp field, graphite hollow cathode discharge has been seen to provide clustering environment conducive to the formation of large carbon clusters $C_m^+ (m \leq 10^4)$ [1]. Our earlier results [2] identified the existence of sooting layers on cathode's inner walls as a pre-condition for the formation of such large clusters. In another study [3] we looked at the photoemission and the velocity spectra of the sputtered negative ions and non-regenerative graphite surface. This communication is an extension of the study of the regenerative soot formation by identifying the state of excitation and ionization of the carbonaceous plasma. Our indicators are the relative number densities of the excited and ionized neon and carbon. Carbon cluster formation in sooting environments has led to the discovery of fullerenes in the laser ablated graphite plumes [4] and also during arc discharges between graphite electrodes [5].

In all of our earlier results [1, 2] we utilized an $E \times B$ velocity filter for the mass spectrometry of the carbon clusters. On the basis of the analyses of these spectra we identified the transition from a pure sputtering mode to a sooting one [2]. In the present paper our aim is to characterize the state of the carbon vapour during this transformation by using photoemission spectroscopy of the excited and ionized plasma species. The ionized component of the plasma is crucial for the sustenance of the discharge. In addition, we discuss the role of the excited species in the re- cycling and regeneration of the cathode deposited clusters. We explore the roles played by the discharge parameters like the discharge voltage $V_{dis}$, discharge current $i_{dis}$ and the support gas pressure $P_g$. Krätschmer *et al.* [5] had also identified the gas pressure as a critical parameter for sooting in the arc discharge between graphite electrodes. We evaluate the nature of the carbon vapour and the sources of the regeneration of soot by using the levels of excitation of the atomic and ionic species, the spontaneous emission *versus* the electron collisional de-excitations, electron temperature $T_e$ and density $n_e$.

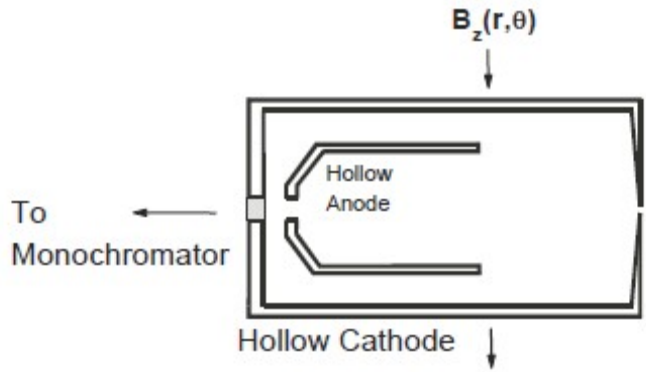

**Fig. 1.** The schematic diagram of the source is shown with graphite Hollow Cathode and Hollow Anode. The cusp magnetic field $B_z(r; \theta)$ is also indicated with arrows.

## 2 Experimental

We present results from the analysis of the characteristic lines' emission from a Ne discharge in a graphite hollow cathode. A hexapole, cusp magnetic field around the tube is essential to initiate and sustain the discharge at moderate discharge voltages and currents. The source is described in reference [1]. In Figure 1 a schematic diagram of the source is shown. Once the Ne plasma is generated, the ionized species sputter the graphite and the carbon components, atomic as well as clusters, are introduced into the plasma. The C content of the plasma reaches a saturation after prolonged operation and we notice a gradual transition from a sputter eroded graphite to the regeneration of the sooted cathode [2]. The figure shows the carbon cluster source with the inter-penetrating Hollow cathode and Hollow Anode. It shows the photoemission spectroscopy being done with a monochromator with 1 Å resolution. Photoemission spectroscopy was done with a compact Jobin Yvon monochromator type H20UV with a grating blazed at 300 nm. The stepper motor was operated with a minimum of 0.1 nm steps. The photomultiplier tube and the grating efficiencies vary between our chosen range of wavelength 180−650 nm. Fused silica window was fitted on the hollow cathode source for the transmission of wavelengths down to 180 nm. The emission spectra are presented with intensities as obtained from the photomultiplier but while calculating the number densities of the excited levels, we have multiplied with the appropriate correction factors for the respective wavelengths.

# 3 Results and discussion

## *3.1 The excited and ionized C content*

The graphite hollow cathode discharge with Ne as the source gas shows three distinct groups of emission lines between 180 and 650 nm. The first group is between 180–250 nm. This includes emission lines belonging to the neutral, singly and doubly charged C denoted as CI, CII and CIII, respectively.

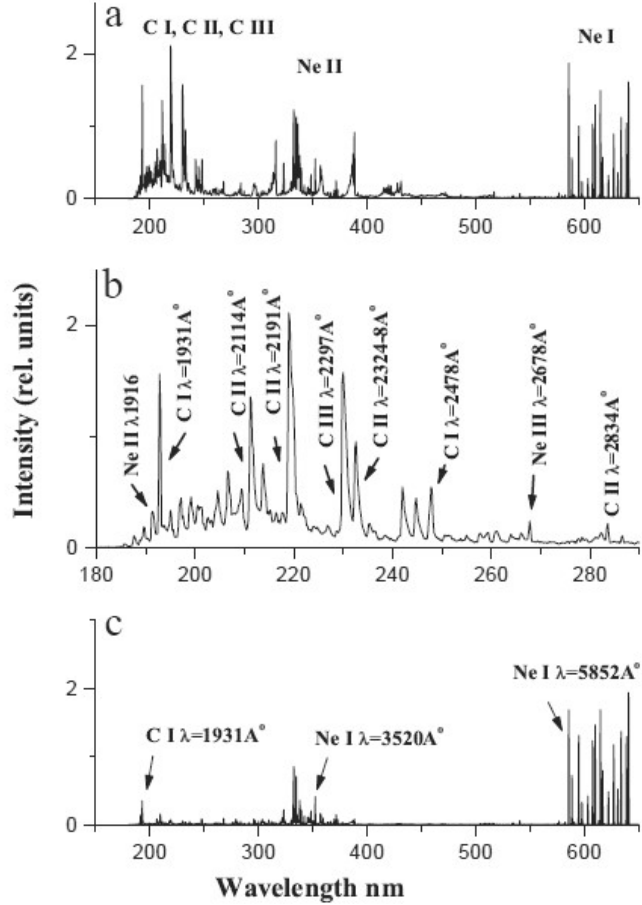

**Fig. 2.** Photoemission spectra of the graphite hollow cathode discharge are shown at $P_{Ne} \approx 0.6$ mbar. Spectrum in (a) is from a fresh cathode surface with $V_{dis}$ = 1.2 kV, $i_{dis}$ = 200 mA. In (b), an expanded version of the 180−290 nm range of (a) is shown to identify the CI, CII and CIII lines; (c) is from a well sooted cathode with $V_{dis}$ = 600 V, $i_{dis}$ = 50 mA.

The lines belonging to NeII for ex- ample at $\lambda = 1916$ Å are also present but this wavelength regime is dominated by carbon's excited atomic and ionized lines. The presence of these lines implies that the initially pure Ne discharge has been transformed into a carbonaceous one. Between 300–400 nm, neon's ionic lines are grouped together with the molecular bands at 357 nm and 387 nm. These will discussed with some other molecular features in Section 3.6. A significant exception is a NeI line at $\lambda = 3520$ Å which is a persistent feature of all the spectra. This is a resonant line of NeI and we use it for the determination of the electron temperature $T_e$. The third distinct and high intensity group of emission lines lies between 580–650 nm and these are exclusively due to the excited atomic NeI levels de-exciting to 3s[1/2,3/2] levels. A large percentage of the discharge power is concentrated in these excited atoms of Ne that cannot de-excite to ground. These excited atoms give up their energy (~16.7 eV per NeI) in collisions with the cathode walls. We have recently explored their soot regenerative properties as a potential sputtering agent [6].

Figure 2 shows three spectra with the source gas pressure $P_{Ne} \approx 0.6$ mbar. We have interpreted the emission lines using NIST's extensive Atomic Spectra Database (ADS) available on the web [7]. During the first spectrum in Figure 2a with $i_{dis} = 200$ mA, the discharge voltage $V_{dis} = 1.2$ kV. This spectrum is taken on a freshly pre-pared cathode surface, at high values of $V_{dis}$ and $i_{dis}$. This is a typical carbon cathode sputtering dominated spectrum. All the respective neon lines both atomic and ionic are present. But the VUV part of the spectrum is dominated by carbon's neutral, singly and doubly charged CI, CII and CIII lines. An expanded view of the wavelength regime between 180–290 nm is shown in Figure 2b. The emission lines $\lambda = 1931$ Å and $\lambda = 2478$ Å are the two dominant VUV lines of CI that can be seen along with the $\lambda = 2135$ Å and $\lambda = 2191$ Å of CII and the $\lambda = 2297$ Å of CIII. Also present is CII's 232 nm inter-combination multiplet. It has very small transition probabilities $\sim 0.1$ s$^{-1}$ for all five lines of the multiplet. This multiplet is generally a weak intersystem transition route for the de-excitation of CII in diffuse interstellar clouds [8]. From the energy level scheme of C in NIST Database [7] we can see that out of the total of 254 CII emission lines between 0–2000 Å, 73 lines are emitted by the de-excitation to the first excited level $2s2p^2[^4P_{1/2}...]$ of CII. This level, in turn de-excites to the ground level $^2P_{1/2,3/2}$ by the emission of the 232 nm multiplet. The intense emission indicates that CII exists as a highly excited C ion in the discharge. In the spectrum shown in Figure 2c, the carbon lines reduce in intensity as the discharge is operated with $V_{dis} = 600$ V and at lower $i_{dis}$. A general trend for the diminishing line intensity of the carbon species is visible. Neon's ionic lines' contributions in the 300{400 nm range also have reduced intensities. These NeI lines between 580-650 nm remain as the significant emission feature of the low $i_{dis}$ spectra.

Singly ionized carbon's first excited state $^4P_{1/2,3/2,5/2}$ has lifetime $\tau(^4P) \approx 4.7$ms. Thus its level density serves as a useful indicator of the carbon content of the cusp field, graphite hollow cathode plasma. In Table 1 we have presented the calculated relative densities from the line intensities of CI $\lambda = 1931$ Å, CII $\lambda = 2324-2328$ Å, NeI $\lambda = 5852$ Å, NeII $\lambda = 3713$Å. Their emission lifetimes are $\tau(\lambda=1931Å)=2.85$ ns, $\tau(233$ nm$)=4.7$ ms, $\tau(\lambda=5952Å) =14.7$ ns and $\tau$ ($\lambda$=3713 Å) = 7.7 ns. From the natural radiative lifetimes of these four excited states CII has six orders of magnitude longer residence time in the plasma.

Table 1. The calculated relative densities NeI ($D_{NeI}$), NeII ($D_{NeII}$), CI ($D_{CI}$), and CII ($D_{CII}$) are tabulated for the discharge current $i_{dis}$ = 50−200 mA at $P_{Ne} \approx$ 0.6 mbar.

|  | 50 mA | 75 mA | 100 mA | 150 mA | 200 mA |
| --- | --- | --- | --- | --- | --- |
| $D_{NeI}$ | 0.44 × 10$^{-7}$ | 0.83 × 10$^{-7}$ | 0.9 × 10$^{-7}$ | 2.14 × 10$^{-7}$ | 2.82 × 10$^{-7}$ |
| $D_{NeII}$ | 0.58 × 10$^{-10}$ | 0.9 × 10$^{-10}$ | 1.45 × 10$^{-10}$ | 2.3 × 10$^{-10}$ | 3.68 × 10$^{-10}$ |
| $D_{CI}$ | 3.3 × 10$^{-10}$ | 6.7 × 10$^{-10}$ | 7.7 × 10$^{-10}$ | 8.7 × 10$^{-10}$ | 21.6 × 10$^{-10}$ |
| $D_{CII}$ | 2.4 × 10$^{-5}$ | 2.9 × 10$^{-5}$ | 2.3 × 10$^{-5}$ | 3.0 × 10$^{-5}$ | 7.0 × 10$^{-5}$ |

We will discuss in the next section that its collisions with the walls are most likely as opposed to the short lived constituents. From Table 1, in the $i_{dis}$ range 50-200 mA we get the ratio of the densities $D_{CII}=D_{CI} \sim (3.5 \pm 0.5) \times 10^4$. Similarly, $D_{NeII}=D_{NeI} \sim (1.3 \pm 0.3) \times 10^{-3}$ and $D_{CI}=D_{NeI} \sim (0.6 \pm 0.1) 10^{-2}$. These results identify a carbonaceous plasma with the ionized C whose ratio with the excited Ne is $D_{CII}=D_{NeI} \sim 2 \times 10^2$. These results also imply that the density of the singly ionized neon $D_{NeII}$ is only $\sim 10^{-5} \times D_{CII}$ throughout our discharge current range. Therefore, our plasma is dominated by the ionized C and has a 2-4% excited Ne.

## 3.2 Spontaneous emission versus the collisional transitions

Three lines from NeI and two from CI are chosen to establish the spontaneous emission as opposed to the electron collisional de-excitations for the source gas – Ne and the sputtered species – C in Figure 3. The 3p' [1/2] level of NeI at 18.72 eV de-excites to three lower levels emitting lines with wavelengths $\lambda$=5882Å, $\lambda$=6030Å and $\lambda$=6164Å. In Figure 3a the number densities of the upper level i.e., 3p' [1/2] are plotted by using $I_{mn} = N_m h\nu_{mn} A_{mn}$, where $I_{mn}$ is the line's intensity, $h\nu_{mn}$ the energy difference between the levels and $A_{mn}$ its Einstein transition probability for spontaneous emission.

The relative number densities are plotted as a function of the discharge current for the three lines. All values are taken from spectra at the same discharge conditions with the variable parameter being $i_{dis}$. In Figure 3b, we have the two carbon lines at $\lambda$ = 1931Å and $\lambda$=2478Å plotted for the same $i_{dis}$ range. These lines originate from the de-excitation of the 3$^1P_1$ level as shown in the inset of the figure. All densities are within 10% of each other at a given $i_{dis}$. From both of the figures the number densities of the respective lines belonging to the two discharge species show that the process of de-excitation of the excited plasma species is dominantly that of the spontaneous emission as opposed to the electron collision induced de-excitation. We will further elaborate this point in the next section with the calculated values of $\tau_{spon}$ *versus* $\tau_{coll}$, the characteristic times for the two respective processes.

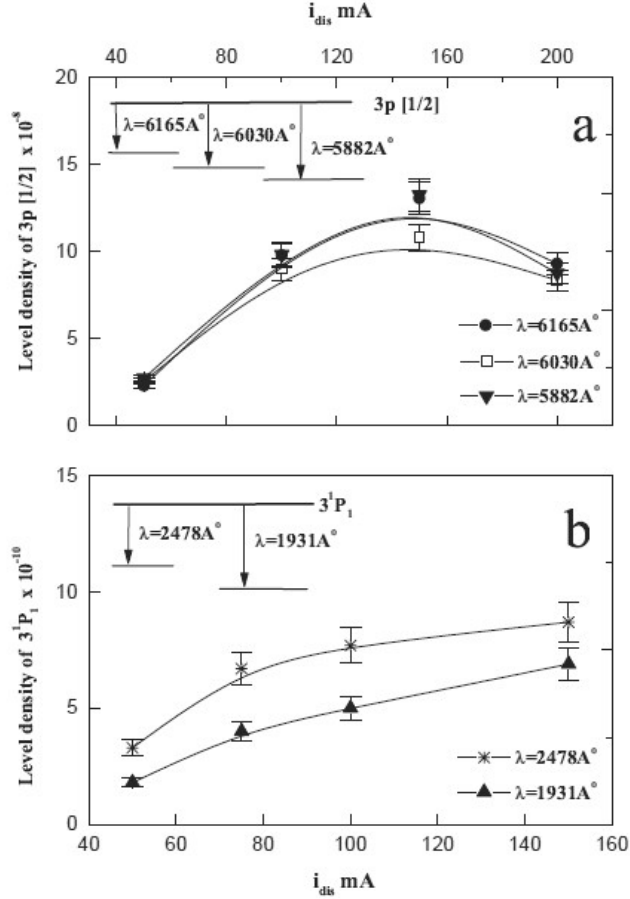

**Fig. 3.** The number densities of the excited levels for NeI denoted as $D_{NeII}$ and CI ($D_{CI}$) are plotted as a function of the discharge current $i_{dis}$ in (a) and (b), respectively. The insets show the upper levels NeI $3p[1/2]$ and CI $3\ ^1P_1$ and the respective de-excitations.

## 3.3 Estimates of $T_e$ and $n_e$

Evaluation of the electron temperature $T_e$ of the plasma from the relative intensities of the spectral lines belonging to the same species can best be done by choosing two lines with transitions to the same lower level. Holtgreven [9] has emphasized the need for at least 1-2 eV energy difference of the two upper terms for higher accuracy in the determination of $T_e$. The excitations in the carbonaceous plasma's atomic and ionic species are induced by electron collisions. Assuming Maxwellian velocity distribution for the electrons, the electron temperature $T_e$ is evaluated from the two resonant lines of NeI $\lambda = 5852$ Å and $\lambda = 3520$ Å by using the Boltzmann equation. The level $3s[1/2]$ at 16.85 eV is populated by the spontaneous emission of these two lines from $3p[1/2]$ and $4p[1/2]$ at 18.96 and 20.37 eV, respectively. This provides $T_e \approx 8500 \pm 300$ K for the discharge current $i_{dis}$ in the range 50-200 mA. If we use 2 non-resonant transitions NeI $\lambda = 3501$ Å and NeI $\lambda = 5400$ Å then we get $T_e \approx 6200 \pm 500$ K for the same $i_{dis}$ range. The excitation temperature for the two sets of the singly charged neon lines NeII is a factor of 2 higher than that obtained by using NeI lines. With NeII $\lambda = 1938$ Å and NeII $\lambda = 3727$ Å we get $T_e \approx 15800 \pm 1000$ K while NeII $\lambda = 1916$ Å and NeII $\lambda = 3717$ Å yield $T_e \approx 18500 \pm 1000$ K. Both sets of these transitions are, respectively, to the same lower levels.

We choose $T_e \approx 8500 \pm 300$ K as the representative electron temperature throughout the discharge current range 50 mA $\leq i_{dis} \leq$ 200 mA for the carbonaceous plasma that we are describing due to the following considerations that:

(1) There is a remarkable consistency in the profile and relative intensity ratio of the resonant emission lines NeI $\lambda = 5852$ Å and $\lambda = 3520$ Å.

(2) The ratio of the densities of the excited carbon to neon $D_{CI}=D_{NeI} \approx (0.6 \pm 0.15) \times 10^{-2}$.

(3) The measured ratio of the singly ionized to the neutral neon $D_{NeII}=D_{NeI} \approx (1.3 \pm 0.3) \times 10^{-3}$.

(4) Using the relative intensity of the CII 232 nm intercombination multiplet, we obtain its ratio to NeI as $D_{CII}=D_{NeI} \sim 10^2$.

With $D_{CII} \sim 10^5 \times D_{NeII}$ the most dominant ionized species in the discharge is CII but the resolution of our monochromator does not allow us to distinguish the individual lines of the 233 nm multiplet that has been used to determine $T_e$ and $n_e$ of astrophysical objects [10]. The Coronal model gives up to 3 times higher values for the electron temperature. The validity of the Coronal *versus* the LTE based calculations depend on the electron density being small *i.e.* $n_e \leq 10^8$ cm$^{-3}$.

Saha's equation is used to estimate the electron density $n_e$ by using the relative density of CII and CI and compared with those from NeII $\lambda = 3713$ Å and NeI $\lambda = 5852$ Å. This estimation of $n_e$ is necessary to establish the roles played by the various discharge species that include all the prominent ionized as well as the excited ones. For the constant pressure $P_{Ne} \approx 0.6$ mbar and $i_{dis}$ between 50 and 200 mA as shown in Table 1, we obtain $n_e$ _ 2 _ 1010 cm−3. This value is obtained from the $D_{CII}=D_{CI}$ ratio at $T_e$ = 8 500 K. By using $D$NeII=$D$NeI ratio a two orders of magnitude higher $n_e \approx 10^{12}$ cm$^{-3}$ is obtained. From these two values we prefer to use the one obtained from $D_{CII}=D_{CI}$ ratio.

$T_e$ and $n_e$ are obtained for those discharge conditions where the cathode surface is freshly prepared *i.e.*, no prior sputtering takes place and one gets the emissions from an unsooted surface as a function of the pressure. The three values of $T_e \approx 16300$ K ($P_{Ne}$ = 0.06 mbar), 10700 K ($P_{Ne}$ = 0.1 mbar), and 8500 K ($P_{Ne}$ = 2 mbar), respectively, have $ne \approx 1.5 \times 10^{13}$, $10^{11}$ and $2 \times 10^{10}$ cm$^{-3}$. A factor of two in variation of $T_e$ has the corresponding $n_e$ varying by three orders of magnitude. As opposed to the sputtering dominated cathode emission we can evaluate the same plasma parameters for a well sooted cathode discharge. As discussed in Section 3.1 the number densities are such that $D_{CI} \sim D_{NeII}$ while $D_{CII} \approx 10^2 \times D_{NeI}$ and $D_{CII} \approx 10^5 \times D_{CI}$. Thus for $i_{dis}$ = 50−200 mA range $n_e \approx 2 \times 10^{10}$ cm$^{-3}$. During these experiments $P_{Ne} \approx 0.6$ mbar but similar values have been obtained with $P_{Ne} \approx 0.1$ mbar. The well sooted discharge operates with stable values of $T_e$ and $n_e$. The spontaneous emissions as opposed to the electron collision induced de-excitations can be investigated by using Regemorter's empirical formula [11]. For $T_e \approx 8500$ K and $n_e \approx 10^{10}$ cm$^{-3}$, we obtain for characteristic times $\tau_{coll} \sim 10^{-3}$ s $\gg \tau_{spon} \sim (10^{-7}-10^{-8})$ s. The result is valid for NeI $3p'$ [1/2], CI $3^1P_1$ and NeII $^2D_0$ levels. Therefore, we are justified in our assumption for the assumption of the spontaneous decay for CI, NeI and NeII. The situation is quite different for CII ($^4P_{1/2,3/2,5/2}$) whose lifetime is in ms. We expect it to participate in the kinetic sputtering of the cathodes along with the potential sputtering by NeI.

Table 2. Ionization rate coefficient $\alpha_i$ cm$^3$ s$^{-1}$ for the atomic and ionic species of C and Ne, CI–CII, CII–CIII and NeI–NeII, NeII–NeIII. All ionizations are from the ground state.

|  | 1 eV | 10 eV | 500 eV |
| --- | --- | --- | --- |
| CI-CII | 1.93 x 10$^{-13}$ | 1.7 x 10$^{-8}$ | 9.7 x 10$^{-8}$ |
| CII-CIII | 1.2 x 10$^{-19}$ | 1.3 x 10$^{-9}$ | 3 x 10$^{-8}$ |
| NeI-NeII | 1.15 x 10$^{-18}$ | 1.32 x 10$^{-9}$ | 6.1 x 10$^{-8}$ |
| NeII-NeIII | 3.25 x 10$^{-28}$ | 9 x 10$^{-11}$ | 2.8 x 10$^{-8}$ |

## 3.4 Two energy regimes of electrons

The role played by the two distinct energy regimes of electrons can provide an estimate of the rather high densities of the ionized species CII, CIII, and NeII, NeIII. These are present even at $i_{dis}$ as low as 50 mA in Figure 2c. Table 2 is prepared by using Lotz' semi empirical formulation [12] for the ionization rate coefficients $\alpha_i$ cm$^3$ s$^{-1}$ for the successively higher ionization stages of C and Ne. For these calculations Maxwellian velocity distribution for the electrons is assumed and all excitations and ionizations are from the ground state. At $T_e \sim 1$ eV or less which is the plasma temperature in our case, the presence of the higher ionized species is much less probable. However, at higher electron energies $E_{elecron} \geq 10$ eV, a significant increase in $\alpha_i$ occurs. Between 10−500 eV energy range, the electrons can ionize all ionic stages of C and Ne with similar orders of magnitude probabilities. A recent spectroscopic characterization of carbon impurities in the tokamak plasmas [13], the spectral line ratios of the C ions have been used to evaluate $T_e$. Estimation of the radiation losses by impurity ions has also been done in this study. They observe CII to CIV in the $T_e \sim 4-40$ eV range. Since in our case the $T_e \sim 1$ eV, the required high energy electrons are provided by the cathode for the ionization for the higher ionization levels of C and Ne. These are available due to the secondary electron emission from the cathode but in much reduced intensities compared with the thermal electrons.

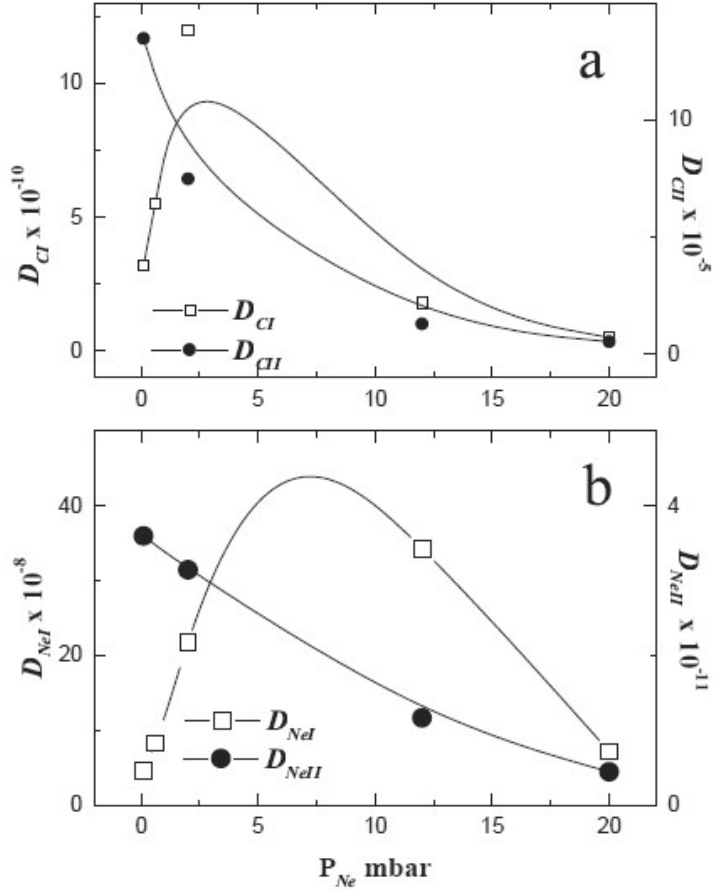

**Fig. 4.** The calculated relative number densities $D_{CI}$, $D_{CII}$, $D_{NeI}$ and $D_{NeII}$ are plotted as a function of Ne pressure.

## 3.5 Indicators of the transition from sputtering to sooting plasma

Kratschmer *et al.* [5] had noticed during their pioneering work on the discovery of fullerene formation that at higher gas pressures their arc discharge with graphite electrodes yielded soot that showed pronounced absorption in the UV range at 215, 265 and 340 nm. These are the characteristic absorptions that characterize the fullerenes. Our experiment is in some respects similar to theirs but with certain differences. Firstly, we produce a regenerative sooting discharge in which the entire soot assembled on the cathode is recycled. Secondly, we monitor the emissions by the excited species as opposed to absorption spectroscopy. We have also varied the gas pressure to see the effects on the carbon content of the sooting plasma and to be able to establish the indicators for the transition from a pure sputtering dominated discharge to sooting plasma that has been seen to produce large clusters as noted in reference [2]. In Figure 4 we have plotted the calculated level densities of $CI(3^1P_1)$,

CII($^4$P$_{1/2;3/2;5/2}$), NeI($3p$'[1/2]) and NeII($^2$D$^0$) as a function of the gas pressure between 0.1 and 20 mbar. Almost five orders of magnitude difference in the level densities of CI and CII is evident in Figure 4a. $D_{NeI}$ and $D_{NeII}$ also show similar pattern in Figure 4b. The pattern of the rise and fall for the excited species as opposed to a gradual downward trend for the ionized ones is also clearly depicted in these figures. $D_{CI}$ shows a peak around 2 mbar while $D_{NeI}$ has it around 7-8 mbar. Unlike the ratios of the results presented in Table 1 where $D_{CII}/D_{CI}$ and $D_{NeII}/D_{NeI}$ were constant as a function of the discharge current $i_{dis}$, we have a gradually decreasing pattern for the two ratios as a function of $P_{Ne}$.

## *3.6 The molecular emission bands*

The other most notable features in the photoemission spectra of Figure 2 are the Swan band heads of $C_2$ at $\lambda$ =5165 Å, $\lambda = $ 4737 Å, $\lambda = $ 4715 Å, and $\lambda = $ 6535 Å. These band heads have recently been discussed elsewhere [14] to identify the parameters for the formation of the $C_1$ and $C_2$ in the hollow cathode discharges. We found that the number density of $C_2$ is an increasing function of $i_{dis}$ at constant pressures. Another band structure at 387 nm and 357 nm have not been fully identified as yet by us. In our experiments we have made elaborate attempts to maintain good vacuum and to avoid molecular gases ($H_2$, $O_2$, $N_2$...) in the discharge chamber. Therefore, the bands at 387 nm and 357 nm are expected to be related with the overall process of the carbon cluster formation and fragmentation. Further photoemission work is in progress to fully understand the state of $C_2$ and these unresolved bands. We have seen these bands in the hollow cathode discharges with various noble gases including He, Ne and Xe. Obviously, more work with better resolution monochromator is highly desirable.

# 4 Conclusions

Two modes of the inclusion of carbon into the neon plasma can be clearly identified from the data on the relative number densities of CI and NeI calculated from the intensities in the photoemission spectra. The first mode is the high $T_e$ and $n_e$ regime of the discharge where higher values of $V_{dis}$ and $i_{dis}$ initiate the sputtering of the carbon cathode with the subsequent excitation and ionization of the carbon atoms with high energy electrons emitted from the cathode and accelerated in the cathode fall to energies ~500 eV. The singly ionized C content participates efficiently in the kinetic sputtering of the cathode along with NeII. Such a discharge is characterized with $V_{dis}$ = 1.2 kV and $i_{dis}$ = 200 mA and the resulting emission spectrum was shown in Figure 2. The second mode of discharge can be classified as the sooting mode which may be associated with high pressure discharges where the density of the ionized species $D_{CII}$ and $D_{NeII}$ considerably reduces. A constant but gentle surface erosion by potential sputtering dominates this mode as has been discussed in reference [6]. Kinetic sputtering is also taking place and the process of cathode sputtering involves the two mechanisms together. CI and CII are the integral constituents of all sooting processes and their densities are indicators of the soot formation on the cathode walls. We envisage the sooting mode to imply a loose agglomeration of carbon clusters on the cathode surface being recycled or regenerated by kinetic sputtering with energies ⪆500 eV as well as the collisions of metastable Ne atoms with energies $E \sim 16.7$ eV. In conclusion, we have identified and presented the parameters required for the transition from the sputtering proficient regime to the sooting dominant profile of the regenerative soot. We believe that this in- formation provides us an understanding the dynamical processes that are responsible for the formation of clusters including the fullerenes in the regenerative sooting discharges.

Our present investigation into the role of the state of the carbon vapour in graphite hollow cathode discharge can yield useful information on certain similar aspects of the plasma-wall sputtering from the first wall of the tokamaks. Although we have designed the source to enhance the cathode sputtering, the reverse is the requirements for plasma containment in a tokamak reactor. The conclusions are also valid for the studies of such processes in tokamaks. Our results show that the extended $3D$ cusp magnetic field contours create an ideal environment for the entrapment of the charged particles. The subsequent interactions of these trapped, positively charged species with the sooted cathode walls provide an efficient recycling mechanism for the regeneration of the soot.

## References

1. S. Ahmad, T. Riffat, Nucl. Instrum. Meth. Phys. Res. B 152, 506 (1999).
2. S. Ahmad, Phys. Lett. A 261, 327 (1999); Eur. Phys. J. Appl. Phys. 5, 111 (1999).
3. A. Qayyum, M.N. Akhtar, T. Riffat, S. Ahmad, Appl. Phys. Lett. 75, 4100 (1999).
4. H.W. Kroto, J.R. Heath, S.C.O'Brien, R.F. Curl, R.E. Smalley, Nature 318, 162 (1985).
5. W. Krätschmer, L.D. Lamb, K. Fostiropoulous, D.R. Huffman, Nature 347, 354 (1990).
6. S. Ahmad, M.N. Akhtar, Appl. Phys. Lett. 78, 1499 (2001).
7. NIST Atomic Spectra Database (ADS) Data at http://physics.nist.gov/
8. Th. Henning, F. Salama, Science 282, 2204 (1998).
9. W.L. Holtgreven, in *Plasma Diagnostics* (North Holland Publishing Co., Amsterdam, 1968), pp. 178-182.
10. R.E. Stencel *et al.*, Monthly Notices Roy. Astron. Soc. 196, 47 (1981); D.J. Lennon, P.L. Dufton, A. Hibbert, A.E. Kingston, Ap. J. 292, 200 (1985).
11. H. van Regemorter, Ap. J. 136, 906 (1962); O. Bely, H. van Regemorter, Ann. Rev. Astron. Astrophys. 8, 329 (1970).
12. W. Lotz, Ap. J. 14, 207 (1967).
13. R.C. Isler, R.W. Wood, C.C. Klepper, N.H. Brook, M.E. Fenstermacher, A.W. Leonard, Phys. Plasmas 4, 355 (1997).
14. S. Ahmad, A. Qayyum, M.N. Akhtar, T. Riffat, Nucl. In- strum. Meth. Phys. Res. B 171, 551 (2000).